\documentclass[num-refs]{nbdt-article}
\usepackage{siunitx}
\usepackage[utf8]{inputenc}
\usepackage{amsmath, amssymb}

\def\a{\mathrm{A}}
\def\b{\mathrm{B}}
\def\ab{\mathrm{AB}}
\def\pois{\textit{Poi}}
\def\gam{\textit{Gam}}
\def\bet{\textit{Be}}
\def\disc{\textit{Discrete}}

\def\iid{\sim}
\def\lambdalo{\underline{\lambda}}
\def\lambdahi{\overline{\lambda}}
\def\IBF{B^*}

\papertype{Original Article}
\paperfield{Analysis}

\title{Sensitivity and specificity of a Bayesian single trial analysis for time varying neural signals}

\author[1-3]{Jeff T. Mohl}
\author[1-5]{Valeria C. Caruso}
\author[1,6]{Surya T. Tokdar}
\author[1-4]{Jennifer M. Groh}

\affil[1]{Duke Institute for Brain Sciences, Duke University, Durham, NC, 27708, USA}
\affil[2]{Center for Cognitive Neuroscience, Duke University}
\affil[3]{Department of Neurobiology, Duke University}
\affil[4]{Department of Psychology and Neuroscience, Duke University}
\affil[5]{Center for Human Growth and Development, University of Michigan, Ann Arbor, MI, 48109, USA}
\affil[6]{Department of Statistical Science, Duke University}

\corraddress{Jeff T. Mohl, Department of Neurobiology, Duke University, Durham, NC,27708, USA}
\corremail{jeffrey.mohl@duke.edu}

\fundinginfo{This work was supported by an NDSEG research fellowship to JTM (32 CFR 168a) and by a grant from the NIH to JMG (R01 DC016363)}

\runningauthor{Mohl et al.}

\begin{document}

\maketitle

\begin{abstract}
We recently reported the existence of fluctuations in neural signals that may permit neurons to code multiple simultaneous stimuli sequentially across time \cite{Caruso2018}.  This required deploying a novel statistical approach to permit investigation of neural activity at the scale of individual trials.  Here we present tests using synthetic data to assess the sensitivity and specificity of this analysis.  We fabricated datasets to match each of several potential response patterns derived from single-stimulus response distributions.  In particular, we simulated dual stimulus trial spike counts that reflected fluctuating mixtures of the single stimulus spike counts, stable intermediate averages, single stimulus winner-take-all, or response distributions that were outside the range defined by the single stimulus responses (such as summation or suppression).  We then assessed how well the analysis recovered the correct response pattern as a function of the number of simulated trials and the difference between the simulated responses to each “stimulus” alone.  We found excellent recovery of the mixture, intermediate, and outside categories (>97\% correct), and good recovery of the single/winner-take-all category (>90\% correct) when the number of trials was >20 and the single-stimulus response rates were 50Hz and 20Hz respectively.  Both larger numbers of trials and greater separation between the single stimulus firing rates improved categorization accuracy.  These results provide a benchmark, and guidelines for data collection, for use of this method to investigate coding of multiple items at the individual-trial time scale. 

\end{abstract}

\section{Introduction}
We recently showed that when multiple stimuli are present, some neurons exhibit activity patterns that fluctuate between those evoked by each stimulus alone \cite{Caruso2018}. This dynamic code could allow the representation of all stimuli within the same population of neurons. Such fluctuations may be a widespread phenomenon in the brain, but would be overlooked using conventional analysis methods that investigate mean activity pooled across trials. Of particular interest are cases in which the time-and-trial-pooled responses evoked by multiple stimuli appear to reflect the average of the responses to each stimulus presented in isolation. This phenomenon, known as divisive normalization \cite{Carandini1994},  has been observed in visual brain areas such as V1 and MT \cite{Britten1999} as well as other sensory and cognitive domains \cite{Bao2018, Kozlov2016, Louie2013, Ohshiro2011a, Olsen2010, Reynolds2009}.  However, such responses could either reflect a true averaging of the two stimuli/conditions, producing a consistent stable intermediate level of firing on each trial, or could reflect a dynamic code that flexibly shifts between the individual stimuli across trials.

To evaluate neural responses on a single trial basis, the novel statistical approach introduced in Caruso et al. (2018) characterizes the distribution of spike counts elicited in response to two simultaneous stimuli using Bayesian inference. Here we provide a general assessment of the sensitivity and specificity of that approach by simulating known neural responses as benchmark cases. In particular, we investigate how the analysis performs as we parametrically varied the data sample size (number of trials), the mean firing rate of responses, and the difference between spike counts across conditions. 

We demonstrate that our approach accurately categorizes synthetic neural data into expected categories. The robustness of the results depends heavily on sample size, as well as on firing rate differences between the two single cue conditions. Importantly, the model performs very well under reasonable experimental values (20 trials per condition, 60\% firing rate change between conditions). Finally, we show that that the model gracefully handles datasets that do not exactly match any of the tested hypotheses. These results demonstrate the viability of the analysis method and provide constraints for interpretation of actual neural data.

\section{Experimental Rationale and Procedures}

\subsection{Neural encoding patterns to be assessed}
For simplicity, our approach focused on the case of two simultaneously presented stimuli (dual stimuli) but can be extended to a larger number of stimuli. We consider an experimental setup in which a neuron’s response is recorded in three interleaved conditions: in the presence of a single stimulus “A”, a single stimulus “B”, or both stimuli A and B (“AB”). We considered four possible response distributions to dual stimuli, in relation to the distributions observed when only one stimulus is present (Figure 1).
\begin{enumerate}
    \item Neurons might respond to only one of the stimuli, and do so consistently (i.e respond to the same one) across trials.  One way this could occur would be if only one stimulus is located in a neuron’s receptive field, but it might also apply when both stimuli are in the receptive field (sometimes referred to as a winner-take-all encoding).  We label this possibility “single”.
    \item The responses to dual stimuli might be greater than the maximum or less than the minimum of the single-stimulus responses. This category includes enhancement/summation, as well as suppression of the response to one stimulus by another. We refer to this case as “outside”.
    \item The responses to dual stimuli are a consistent weighted average of the responses evoked by each stimulus alone.  Here, the dual stimulus responses are between the bounds set by the two single stimulus responses, and cluster around a stable intermediate value.  We refer to this case as “intermediate”.
    \item The responses to dual stimuli may fluctuate such that on each trial the neuron appears to be responding to only one of the two stimuli. We term this possibility “mixture” because it reflects a mixture of two distributions of A and B stimulus responses. This is analogous to a winner-take-all except that the neuron is switching across trials rather than encoding the same stimulus each trial.  Like the “intermediate” category, there could be a weighting factor such that a higher proportion of trials favor one stimulus over the other.    
\end{enumerate}

\begin{figure}[!ht]
\centering
\includegraphics[width=0.7\textwidth]{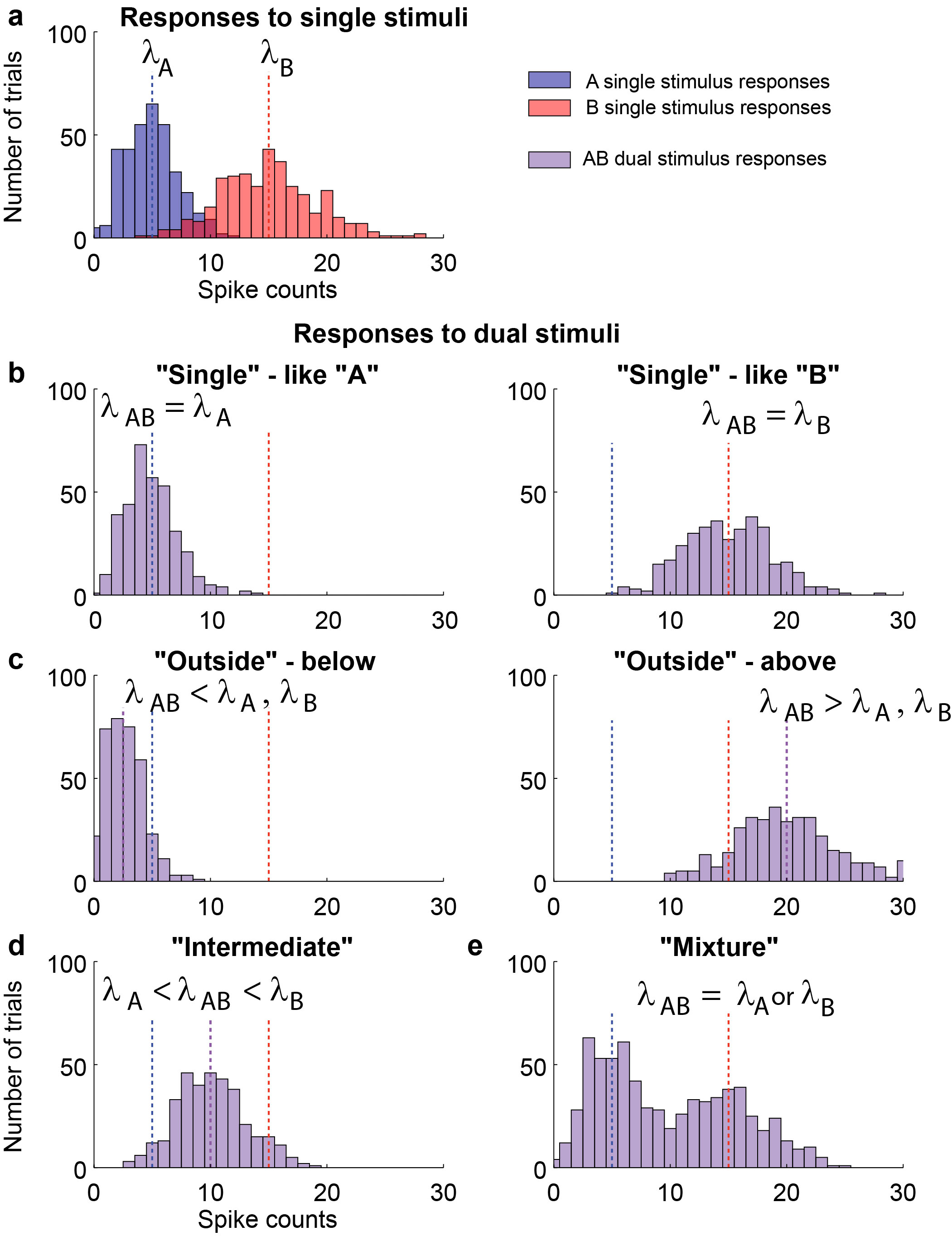}
\caption{Four possible response patterns to dual stimuli trials, in relation to the responses observed to the component stimuli when presented individually.  (a)  Single stimulus trials were modeled as evoking spike counts distributed according to a Poisson process with rates $\lambda_A$, blue, or $\lambda_B$, red.  (b)  Responses on dual stimulus trials follow a Poisson $\lambda_{AB}$ matching either $\lambda_A$, left, or $\lambda_B$, right. (c)  The Poisson rate $\lambda_{AB}$ on dual stimulus trials is less than (left) or greater than (right) those observed in single stimulus trials.  In these simulations, $\lambda_{AB}$ was set to $0.5\cdot\lambda_A$ (left) or $\lambda_A+\lambda_B$ (right).  (d)  Spike counts derived from a Poisson process with a rate $\lambda_{AB}$ between $\lambda_A$ and $\lambda_B$.  (e)  Spike counts drawn from a mixture of two Poissons with rates $\lambda_A$ and $\lambda_B$.}
\end{figure}
\subsection{Model construction, Bayesian model comparison, and synthetic data}

These four possibilities can be formalized on the basis of how the spike distributions on combined stimulus trials AB compare to those observed when only A or B are presented alone.  If A and B elicit spike counts according to Poisson distributions $Poi(\lambda_A )$ and $Poi(\lambda_B)$, then we can ask which of four competing hypotheses best describe the spike counts observed on combined AB trials:

\begin{enumerate}[label=(\alph*)]
\item 	Single: $F=Poi(\lambda)$ for either $\lambda=\lambda_A$ or $\lambda=\lambda_B$, with $\lambda$ constant across trials
\item Outside: $F=Poi(\lambda)$ for some unknown $\lambda\notin[min(\lambda_A,\lambda_B ), max(\lambda_A,\lambda_B)]$
\item Intermediate: $F=Poi(\lambda)$ for some unknown $\lambda\in[min(\lambda_A,\lambda_B ), max(\lambda_A,\lambda_B)]$
\item Mixture: $F = \alpha \cdot Poi(\lambda_A) + (1-\alpha) \cdot Poi(\lambda_B)$ for some unknown $\alpha \in (0,1)$
\end{enumerate}

The plausibility of each of these models was determined by computing the posterior probabilities of each model given the data, with a default Jeffreys’ prior \cite{Berger2006} on each of the model specific rate ($\lambda$) parameters and on the mixing probability parameter ($\alpha$).  Each model was given a uniform prior probability (1/4) and posterior model probabilities were calculated by computation of relevant intrinsic Bayes factors \cite{Berger1996} (see appendix A.1 for a thorough description of the models and model selection strategy).

To evaluate the sensitivity and specificity of this method, we built synthetic neuronal spiking datasets to match each of the four potential encoding strategies tested by the model.  Consistent with our previous study \cite{Caruso2018}, we focused on response patterns that could be modeled as deriving from Poisson distributions.  In principle, the approach could be extended to other forms of response distributions, but this is beyond the scope of this work.

Data files were generated as spike times drawn using an independent Poisson point process sampled at 1 ms intervals, with constant mean firing rate for 1000 ms (Figure 2a-c). For A and B (single stimulus) trials, Poisson rates were assigned a priori to reflect a range of realistic firing rates for a single neuron presented with different stimuli. AB (combined stimulus) trials for each dataset were generated based on the chosen A and B firing rates and in a manner consistent with one of the four potential hypotheses. For the “single” hypothesis the AB data were generated using a single Poisson with rate $\lambda_{AB}$ equal to the highest of the component rates, i.e. $max(\lambda_A,\lambda_B)$. For the “outside” hypothesis, the rate $\lambda_{AB}$ was set 20\% higher than $max(\lambda_A,\lambda_B)$.  For the “intermediate” hypothesis, $\lambda_{AB}$ was equal to the mean of A and B rates $\lambda_{AB} = 0.5(\lambda_A) + 0.5(\lambda_B)$. For the across trial switching "mixture" model, the data were generated using the same Poisson process, but each trial was randomly chosen to be drawn from either $Poi(\lambda_A)$ or $Poi(\lambda_B)$  with equal probability. This results in a dataset for which the across trial average firing rate is equal to the average of the $\lambda_A$ and $\lambda_B$ rates, but individually each trial is better described as deriving from either the $\lambda_A$ or $\lambda_B$ response distributions.  Note that it is nearly impossible to tell by visual inspection of a raster plot when a neuron has such a mixed response pattern, even when the trials are sorted as they are in Figure 2c, but the pattern becomes more evident in histograms of the trial-wise spike counts (Figure 2d).

Multiple datasets were generated using this strategy in order to test the power and reliability of the analysis under plausible experimental conditions. These datasets varied both the number of trials per condition (5-50 trials per condition) and the firing rates of the A and B conditions (from 5-100 Hz, with relative separation between A and B rates of 33-80\% of maximum rate). Individual triplet pairs were generated under each of these conditions, analogous to running 100 individual cells through the analysis. This set of parameters was used for all conditions tested, including datasets constructed to not exactly match any of the hypotheses, discussed in the final section of the results. 

\begin{figure}
\centering
\includegraphics[width = \textwidth]{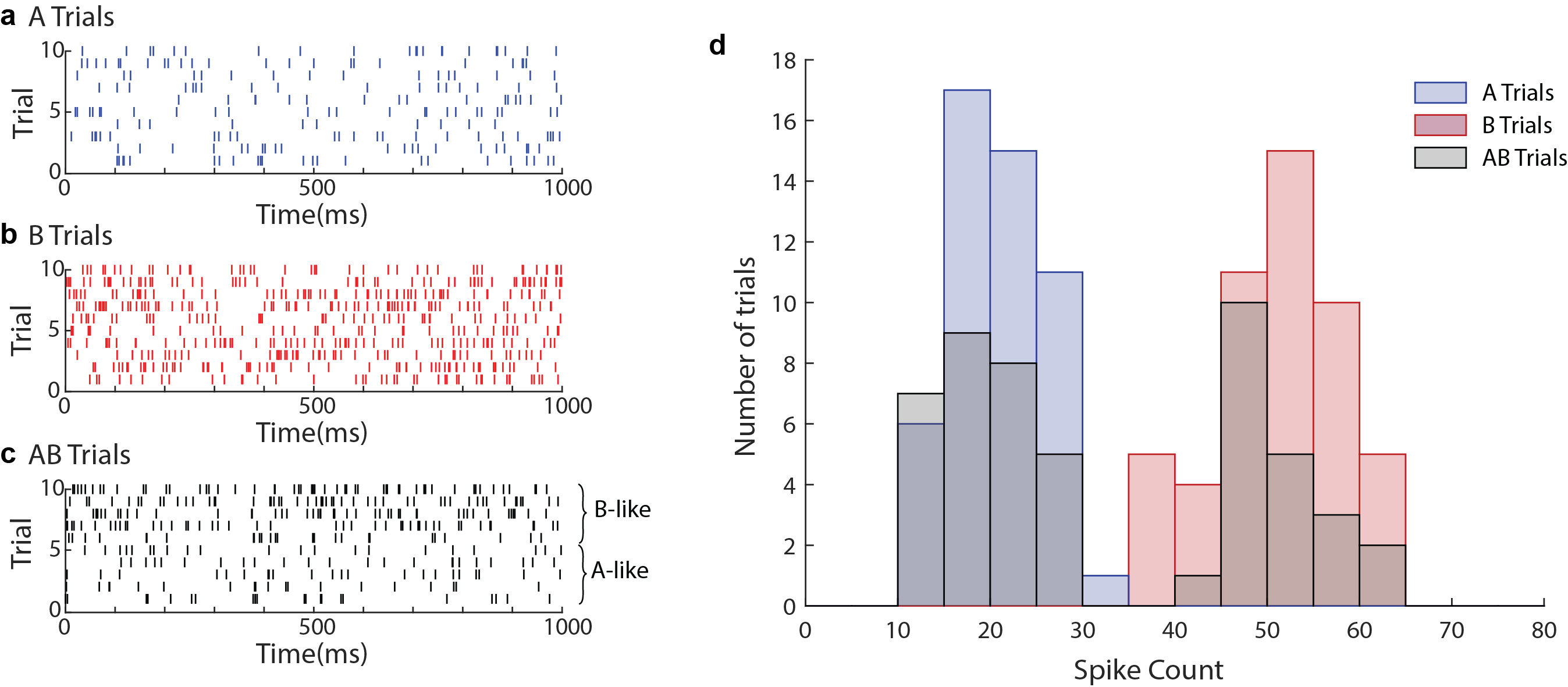}
\caption{One example synthetic dataset. (a-c) Raster plots for a synthetic dataset built to match the across trial switching hypothesis, where blue rows (a) are single A trials, red rows (b) are single B trials, and black rows (c) are AB trials. AB trials are drawn randomly to match either A or B rates and are sorted so that B-like rates are towards the top of the raster. Even with sorting, this pattern is challenging to see with the naked eye, highlighting the need for analytical methods. (d) Whole trial spike count histogram for 50 repetitions of A, B, and AB trials. From this plot, the bimodality of AB trials for the switching condition is more apparent.}
\end{figure}

\subsection{Code and data availability}

Code specific to this paper can be found on GitHub at \href{url}{https://github.com/jmohl/mplx\textunderscore{}tests}, (archived DOI: 10.5281//zenodo.3508536) which includes the code used to generate synthetic data for this manuscript as well as all code needed to perform the neural mixture analysis. The exact synthetic data files used to generate plots are available upon request. Source code and documentation for the Neural Mixture Model available at \href{url}{https://github.com/tokdarstat/Neural-Multiplexing}.

\section{Results}

\subsection{Neural Mixture Model accurately characterizes synthetic data}

The desired analysis outcome is for the output to match the input.  That is, data explicitly generated to match the single hypothesis should be correctly labeled as “single”, data generated as “outside” should be labeled “outside”, etc.  Figure 3 illustrates that this is largely the case.  The series of simulations shown here involved 20 A trials simulated with $\lambda_A = 20$Hz, 20 B trials with $\lambda_B = 50$ Hz, and 20 AB trials generated according to the various methods specified above.  The “mixture” and “intermediate” categories perform exceptionally well, with 100/100 “mixture” and 99/100 “intermediate” datasets labeled correctly with >95\% confidence (dark black bars). This distinction is critical, as these two conditions would produce exactly the same mean rate when averaging across trials, making them indistinguishable using typical neural analysis strategies which average across trials in order to reduce noise. “Single” and “outside” datasets were also correctly labeled in the majority of cases (90/100 and 97/100, respectively), although these hypotheses are not the focus of our analysis as they can be differentiated more easily using simpler statistical methods.  

Although the category “single” was correctly identified as the best model for the dataset simulated under the “single” hypothesis 90\% of the time, the posterior probability or confidence level did not reach the 95\% level observed for the other models.  This is due to the narrow definition of this category:  response rates on AB trials must be indistinguishable from those occurring on either the A or B trials.  All other categories include a range of possibilities which admits this hypothesis as a boundary case (i.e. a weighted average with the weight for A set to 1). Therefore, these models are all competitive in explaining data that is generated to match the “single” case, which explains the low posterior probability of this model. For this reason, it is better to consider the “single” category as reflective of a null hypothesis, where there is no interaction at all between stimuli.

\begin{figure}[!ht]
\centering
\includegraphics[width = 0.95\textwidth]{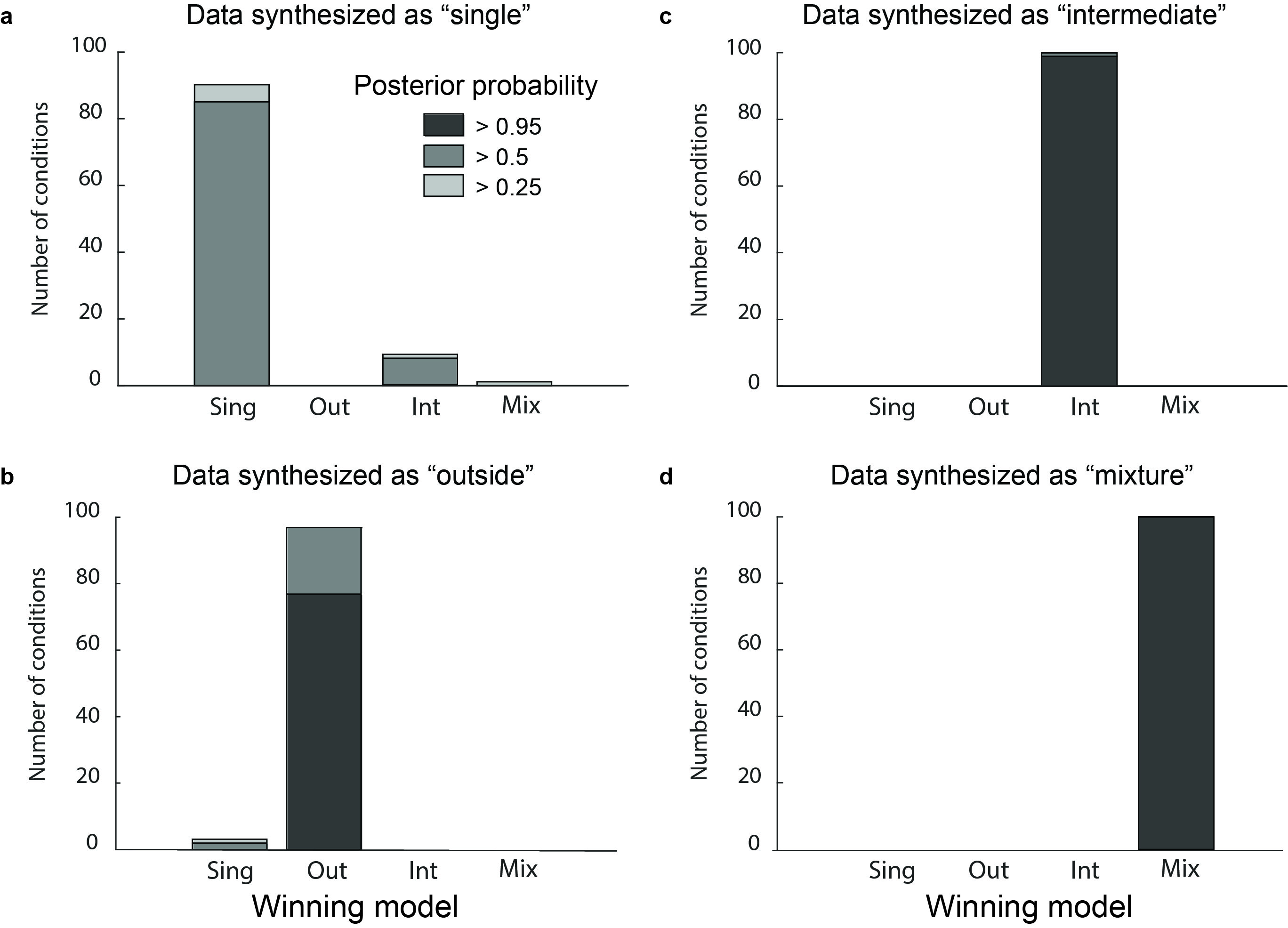}
\caption{The analysis method correctly categorizes synthetic datasets created to match each model. The shading of the bars indicates the posterior probability with which each individual run of synthetic data (n=100) is assigned to a given category. Of particular interest is the very strong separation between intermediate and mixture datasets, as this discrimination is not possible when considering only firing rates averaged across trials. Parameters used for this figure: $\lambda_A = 20$Hz, $\lambda_B=50$ Hz, number of trials in each run = 20 per stimulus condition (60 overall).}
\end{figure}

\subsection{Dependence on number of trials and difference between A and B responses}

The accuracy of this characterization depended on both the amount of data and the difference between the response distributions on A and B trials.  The dependence on the number of trials is best appreciated when considering similar A and B response distributions, such as $\lambda_A = 20$ vs $\lambda_B = 30$Hz shown in Figure 4a, which depicts the average posterior probability value for the correct model as a function of the number of trials.  Even with this modest separation between the A and B response patterns, increasing the number of trials per condition allowed the analysis to better characterize the underlying rates, and therefore better discriminate between competing hypotheses. “Single”, “Intermediate” and “Mixture” had average posterior probability values >0.3 for n=5 trials, but performance improves steadily to average posterior probability values of >0.75 for n=50 trials. When response distributions were moderately separated, $\lambda_A = 20$ vs $\lambda_B = 50Hz$ (the same separation used in Figure 3), performance rose more rapidly for all models except “single”. At n=5, posterior probability values range from 0.4 for “single” to 0.8 for “mixture”.  At n=30, posterior probability values equaled approximately 1 for “mixture”, “intermediate” and “outside”. Further increasing the firing rate separation to $\lambda_A = 20$ vs $\lambda_B = 100$Hz resulted in very high posterior probabilities of >0.95 even at n=5 for “mixture” and “intermediate”; this level was achieved for “outside” at n=10. 

\begin{figure}
\centering
\includegraphics[width = 0.80\textwidth]{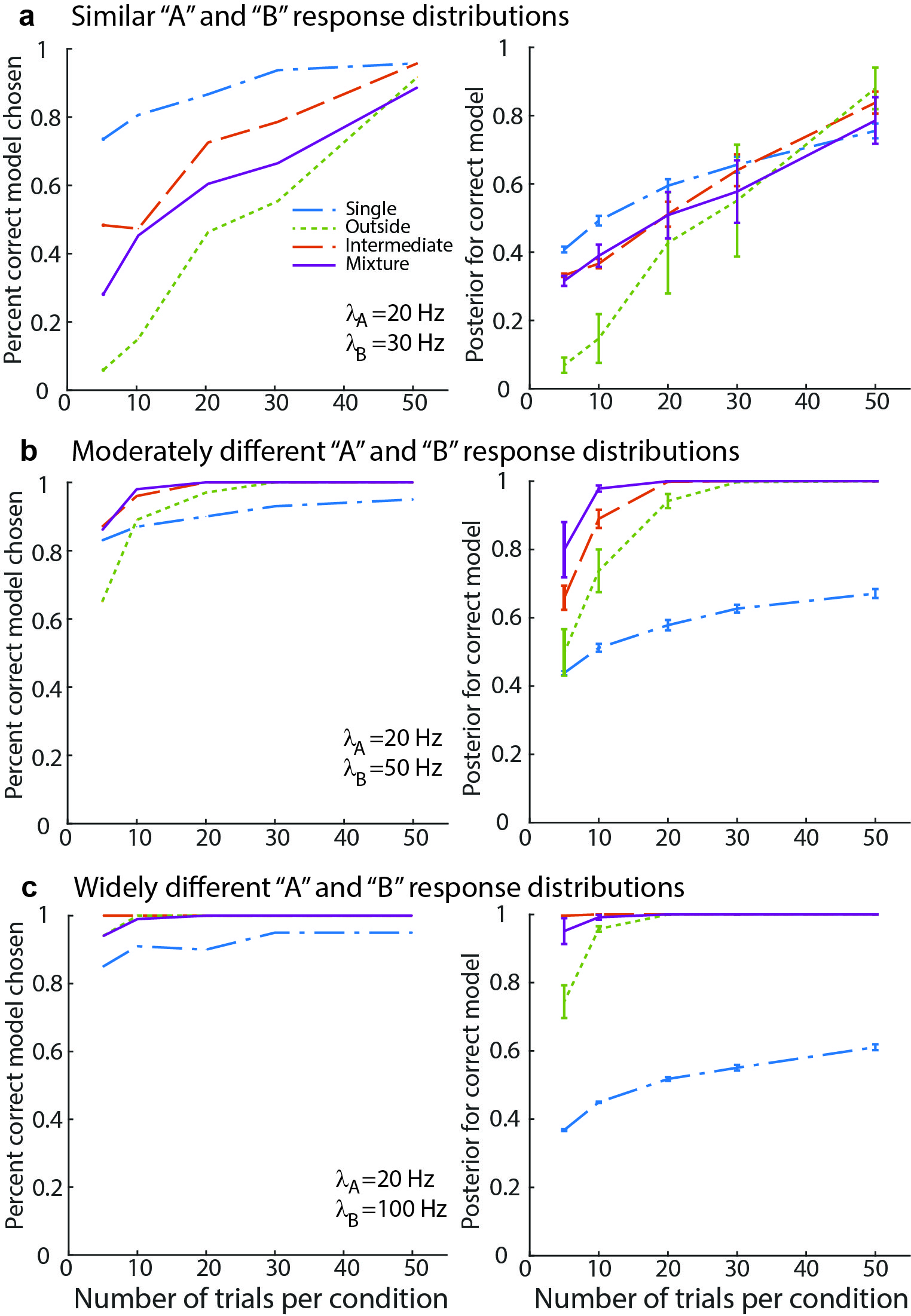}
\caption{Increasing number of trials or separability of conditions improves accuracy of model comparison. (a) Left, percent of triplets which were correctly categorized, split by dataset type, for increasing number of trials per conditions; $\lambda_A = 20$Hz, $\lambda_B=30$Hz. Right, mean and variance of posterior probability for correct model across triplets (b \& c) same as in (a) but with $\lambda_B$ set to 50z and 100 Hz respectively. Fewer trials are needed when responses are very different between A and B trials.}
\end{figure}

These figures give a rough sense of the sensitivity of our analysis, demonstrating that the analysis becomes more reliable as more trials per condition are added until reaching asymptote around 30 trials/condition for a 50 Hz vs 20 Hz comparison (Figure 4b).  Similarly, increasing the difference in spike count between A and B conditions also improves specificity in the analysis, allowing for accurate characterization with as few as 5 trials (Figure 4c). Although these data were constructed under ideal conditions (the data perfectly matches one of the tested hypotheses), they can be used as a guide for how much data should be collected in order to obtain satisfactory results in a real dataset.

The above results give a detailed view of how each model performs across a realistic range of firing rates for neural recordings from primate sensory cortices and sub-cortical areas, for which we initially designed this method. We next sought to determine whether the analysis effectively extended into datasets with much lower maximum firing rates. To address this question we performed the analysis on data for a wide range of maximum firing rate values (5 to 100 Hz) for two fixed amounts of relative separation between A and B rates (40\% and 80\% of maximum rate) and characterized the prediction accuracy under each model (Figure 5).  We found that firing rate affected the accuracy of the model, with lower average firing rate conditions resulting in worse performance than higher firing rates. As expected, a larger relative separation between A and B responses (analogous to having stronger neuronal preference for one or the other condition) resulted in significantly better performance, even for low firing rate conditions. However, even when considering the larger separation value of 80\%, datasets with a maximum firing rate of <15 Hz barely reached 95\% accuracy with 50 trials per condition. These results suggest that datasets with very low average firing rates (less than ~15 Hz for the most preferred response) may not be resolvable using this analysis method.

\begin{figure}
\centering
\includegraphics[width = 0.95\textwidth]{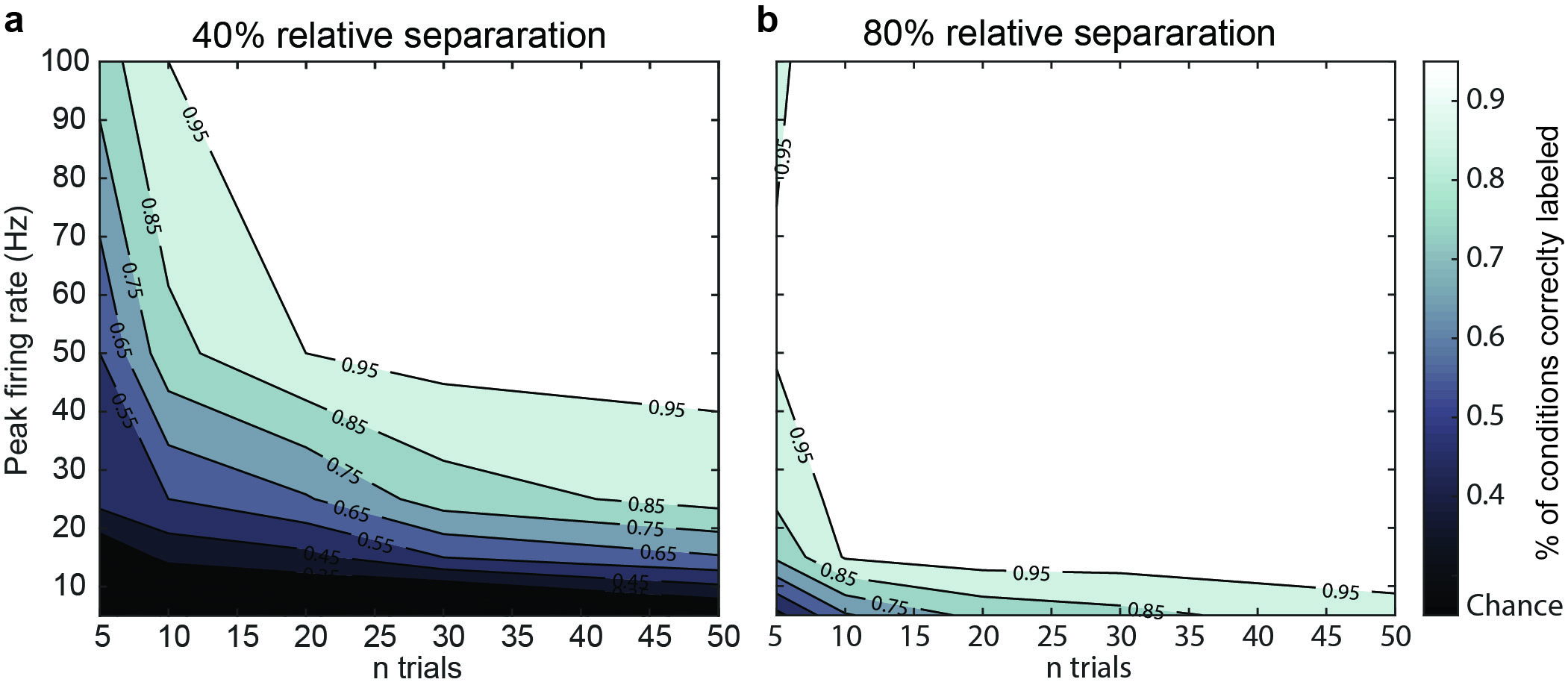}
\caption{Model prediction accuracy depends on both number of trials and average firing rate. (a) Model classification accuracy collapsed across all four models for datasets generated for a fixed relative separation between A and B responses of 40\% of the peak firing rate. Shading shows the percentage of conditions correctly categorized with interpolated phase transitions. (b) Same as in a, but for a relative separation of 80\% of the peak rate.}
\end{figure}

\subsection{Model results are informative when datasets do not perfectly match hypotheses}

Though this modeling strategy was meant to test between discrete hypotheses, it is unlikely that real neural signals perfectly and uniquely match any one of these scenarios. Therefore, it is important that the analysis accurately reflect deviations from exact hypothesis matches. Here, we consider two potential deviations from the circumstances considered above: weighted averaging of A and B stimuli and incomplete switching between A and B rates.

For weighted averaging datasets the AB trials were generated as in the “intermediate” condition above, except that the AB rate was set to be closer to the A rate than the B rate: $\lambda_{AB}=0.75\cdot\lambda_A+0.25\cdot\lambda_B$. Because the model returns both a classification and a posterior probability (reflecting the model’s confidence in that classification), we expected that this type of dataset will result in a spread across single and average classifications, but with lower confidence in this assessment. This is indeed the case, as the analysis returned primarily the intermediate category, with some single winners, but with much lower posterior probabilities than the well matched datasets (Figure 6a, compare with Figure 3c).

\begin{figure}
\centering
\includegraphics[width = 0.95\textwidth]{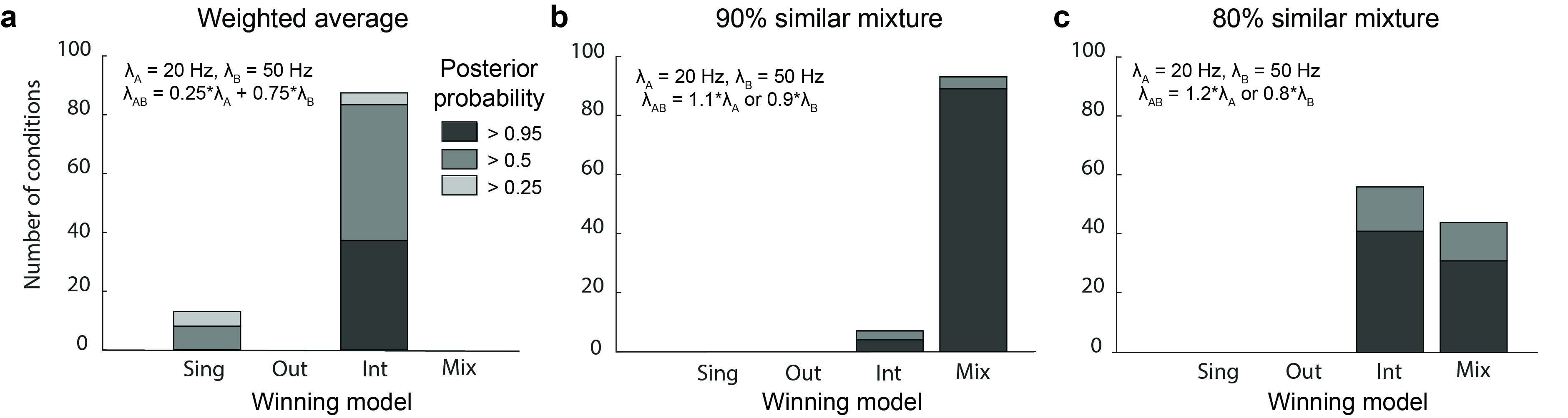}
\caption{Datasets which do not exactly match the canonical hypotheses are descriptively categorized by model comparison. (a) A dataset generated to reflect weighted average of A and B stimuli is weakly categorized as intermediate with some triplets categorized as singles. (b) Mixture trials generated to alternate between values shifted 10\% from the true A and B firing rates are primarily categorized as mixtures. (c) Mixture trials with rates shifted 20\% from the true A and B rates are categorized as either mixture or average with equal probability, consistent with the fact that these trials would be much more difficult to discriminate from the true averaging hypothesis. Parameters used for this figure: $\lambda_B = 50$Hz, $\lambda_A = 20$Hz, number of trials=20.}
\end{figure}

We also tested a form of incomplete switching, where stimuli show strong fluctuations but did not quite exactly match the A and B rates. These datasets were generated using the same strategy as the mixture datasets described above, except that A-like or B-like trials were generated with a slightly shifted mean firing rate. Multiple degrees of similarity were tested, but two (80\% and 90\% similarity) are presented here. From these, the analysis accurately described a 90\% switch as mixture (Figure 6b), but around 80\% similarity it began to interpret many datasets as average (Figure 6c). This highlights a natural limitation that should be expected in the data, as continuing to reduce the similarity would eventually result in a condition that was indistinguishable from true averaging. However, these results demonstrate the high specificity of our analysis for the mixture category, enforcing a strong definition of mixture (literally switching between rates closely matched to the A and B rates, rather than any amount of fluctuation).

\section{Discussion}

There is broad interest in understanding the nature and significance of firing patterns in the brain. It is well known that such firing patterns are variable in the face of identical, highly controlled experimental conditions (such as the presentation of the same stimulus in the same context).  While many studies have viewed this variability as deleterious “noise” that if unsolved would undermine the ability of the brain to perform its essential tasks \cite{Shadlen1994, Shadlen1995, Shadlen1998, Softky1993, Zohary1994}, we and others have sought explanations under the possibility that certain forms of such variation may contribute in a positive fashion to brain function \cite{Caruso2018, Akam2014, Hoppensteadt1998, Jezek2011, Li2016, Lisman1995, Lisman2013, McLelland2016, Meister2002}. In particular, we have successfully modeled variation in whole-trial spike counts to multiple stimuli as being drawn from the observed distributions of spike counts to those same stimuli when presented individually \cite{Caruso2018}.

Here, we benchmarked this analysis on synthetic datasets to provide insight into the sensitivity and specificity of our analysis method as a function of trial counts and the separation between the distributions of spike counts elicited by the component individual stimuli.   When response separations are large, e.g. the mean rate for one stimulus condition is 5X the rate for the other, the analysis method can successfully distinguish among the 4 competing hypotheses with as few as n=5 trials for each condition (n=15 overall). Smaller response separations can be compensated for by collecting more trials to achieve similarly good results.  Finally, even when the conditions do not exactly match the assumptions, such as if the component response rates in the “mixture” condition do not precisely match those observed when the single stimuli were presented individually, correct classifications greatly outnumber incorrect ones.  Critically, the analysis is conservative against the “mixture” hypothesis in these cases, demonstrating that data which is best fit by this model is truly fluctuating between the responses to single stimuli at the single trial level.

These results suggest that the analysis tested here are suitable for many electrophysiological datasets which match the A, B, AB format. Datasets which have moderately high peak firing rates (50 Hz) and an average response difference between conditions of approximately 40\% (relative to peak rate) can reach over 95\% categorization accuracy with as few as 20 trials. Higher peak firing rates, larger separation between neural response, or a greater number of trials all improve the accuracy of our analysis. Conversely, datasets with low peak firing rates (15 Hz) are likely to produce only weak results even with a large number of trials (which will be reflected in low model posterior probabilities). Practically, this means that our analysis is particularly well suited for recordings in primate sensory or motor brain regions with the pronounced tuning and firing rate changes required to differentiate responses at the single trial level.

A situation not tested here is the case in which the response distributions are not derived from Poisson distributions.  In our previous work \cite{Caruso2018} we excluded conditions in which the responses to individual stimuli did not satisfactorily resemble Poisson distributions in order to ensure that our model assumptions were not violated, but this has several downsides.  First, it is difficult to have confidence in the success of this exclusion criterion:  failing to reject the Poisson assumption is not the same as confirming its validity.  Second, a considerable amount of data is excluded in this fashion (as much as 25-50\%, depending on dataset, before even considering other exclusion criteria). Finally, there is significant evidence in the literature that spike counts in many brain areas are more variable than would be suggested by a Poisson distribution \cite{Shadlen1998, Amarasingham2006, Barberini2011, Carandini2004, Cur1997, Recanzone1998}. For all of these reasons, it will be important to both test the model with data sets that violate this assumption and extend the analysis method to include other response distributions such as negative binomials. 

The data presented here reflect conditions where two “stimuli” are presented at the same time, but this analysis could in principle be extended to combinations of three or more response patterns. We have previously shown that responses to multiple auditory stimuli in the primate inferior colliculus are often well described by mixtures of single stimulus responses \cite{Caruso2018}, but little is known about how this type of code changes as more stimuli are added. An extension of this analysis into more complex mixtures of multiple different responses may help bring clarity to this question, and more work is needed to determine whether this phenomenon is general or limited to two stimulus cases.  

Given the broad interest in both noise as a potential limitation on neural representations and in divisive normalization as an elemental computation in sensory processing – with recent suggestions that this process may be impaired in conditions such as autism \cite{Rosenberg2015, Rosenberg2019, VandeCruys2018} – it will be increasingly important to develop additional methods which can probe neural codes at the individual trial level \cite{Macke2015, Pandarinath2018, Park2014}. The tools described in the present paper represent an important step towards uncovering fluctuating patterns in neural activity that may permit greater amounts of information to be encoded in the spike trains of individual and populations of neurons.

\section*{Acknowledgements}
We would like to thank Shawn Willett and Meredith Schmell for help on an earlier version of this manuscript.

\section*{Conflict of interest}
The authors have declared no competing interests related to this work.

\bibliography{mplx-vetting}

\newpage
\appendix

\section{Supplemental Methods - Model Details}
\subsection{Introduction}

Here we record in detail the modeling strategy used and specifics of the model selection procedure, the results of which are reported in the main text. 

\subsection{Model}
For each experimental condition $e \in \{\a,\b, \ab\}$, let $Y^e_j$, $j = 1, \ldots, n_e$ denote the spike counts from all $n_e$ trials run under the condition. We model
\begin{enumerate}
    \item $Y^\a_j \iid \pois(\lambda_\a)$, $Y^\b_j \iid \pois(\lambda_\b)$ for some unknown $\lambda_\a, \lambda_\b > 0$; and,
    
    \item $Y^\ab_j \iid F$ with four competing hypotheses describing $F$
    \begin{enumerate}
        \item Mixture: $F = \alpha \cdot \pois(\lambda_\a) + (1 - \alpha) \cdot \pois(\lambda_\b)$ for some unknown $\alpha \in (0, 1)$
        
        \item Intermediate: $F = \pois(\lambda)$ for some unknown $\lambda \in (\min(\lambda_\a, \lambda_\b), \max(\lambda_\a, \lambda_\b))$
        
        \item Outside: $F = \pois(\lambda)$ for some unknown $\lambda \not\in [\min(\lambda_\a, \lambda_\b), \max(\lambda_\a, \lambda_\b)]$
        
        \item Single: $F = \pois(\lambda)$ for either $\lambda = \lambda_\a$ or $\lambda = \lambda_\b$, with the exact situation being unknown {\it a priori}.
    \end{enumerate}
\end{enumerate}

\subsection{Method}
\subsubsection{Bayesian testing}
We carry out statistical testing between the set of hypotheses $H$ listed above by adopting a Bayesian perspective. A prior probability $p_h$ is assigned to each hypothesis $h \in H$, with $p_h > 0$ and $\sum_{h \in H} p_h = 1$. Let the observed data be denoted $Y = (Y^e_j: 1 \le j \le n_e, e \in \{\a, \b, \ab\})$. Each hypothesis $h$ gives rise to a model for $Y$ which can be written generically as 
\[
Y \sim f_h(y \mid \theta_h), \theta_h \in \Theta_h
\]
where $\theta_h$ captures all unknown parameters under the hypothesis. The modeling process is completed by assuming a prior distribution $\pi_h(\theta_h)$ on the parameter space $\Theta_h$ to reflect prior information and beliefs about the uncertainty about $\theta_h$. Inference about $\theta_h$ is then drawn based on the uncertainty quantified by the resulting posterior distribution $\pi_h(\theta_h \mid Y) = \pi_h(\theta_h) f_h(Y \mid \theta_h) / f_h(Y)$ over $\Theta_h$ where the normalizing constant
\[
f_h(Y) := \int_{\Theta_h} f_h(Y \mid \theta_h) \pi_h(\theta_h) d\theta_h
\]
is recognized as the {\it marginal likelihood score} for hypothesis $h$ given the observed data. Inference about the relative merits of the competing hypotheses is then drawn based on the posterior hypothesis probabilities 
\begin{equation}
p_h(Y) = \frac{p_h f_h(Y)}{\sum_{h' \in H} p_{h'} f_{h'}(Y)}, h \in H,
\label{eq:ph}
\end{equation}
which capture the post-data certainties about the competing hypotheses.

\subsubsection{Prior specification}
Because the four competing hypotheses are only about the distribution of AB trial counts, and do not differ in their description of A and B trial count distributions, we adopt a common prior for the parameters pertaining to these latter distributions. Specifically, we take $\lambda_\a \sim \gam(a, b)$ and $\lambda_\b \sim \gam(a, b)$ for each of the four models. 

For the Mixture hypothesis, the remaining model parameter is the mixing proportion $\alpha \in (0,1)$. We assign it a beta prior: $\alpha \sim \bet(c_1, c_2)$. For the Intermediate hypothesis, given $\lambda_\a$ and $\lambda_\b$, the remaining parameter $\lambda$ is assigned a conditional gamma prior $\gam(a,b)$ truncated to the interval $(\min(\lambda_\a, \lambda_\b), \max(\lambda_\a, \lambda_\b))$. Similarly, for the Outside hypothesis, we take the conditional prior on $\lambda$ given $\lambda_\a, \lambda_\b$ as the $\gam(a, b)$ distribution truncated to $(0,\infty) \cap [\min(\lambda_\a, \lambda_\b), \max(\lambda_\a, \lambda_b)]^c$. In both these cases, the same $a, b$ values are used as for the prior distributions for $\lambda_\a$ and $\lambda_\b$.

\subsubsection{Computation}
Computation of marginal likelihood scores $f_h(Y)$ is generally a complex task in Bayesian inference and require customized approaches to numerically evaluate the integration. Our prior choices and the low dimensionality of the parameter spaces associated with all the hypotheses make the task slightly easier for our problem. However, each hypothesis demands a different strategy to perform the integration and we give enough details below so that an enterprising student can implement these strategies from scratch.

Before getting into the details, we note a particular simplification that can be made to the expression of $p_h(Y)$ in \eqref{eq:ph} thanks to the assumption of a common prior distribution on $\lambda_\a$ and $\lambda_\b$ across all four hypotheses. Write $Y = (Y^\a, Y^\b, Y^\ab)$ where each $Y^e$ denotes the data corresponding to experimental condition $e \in \{\a, \b, \ab\}$. Then we can write
\begin{equation}
    p_h(Y) = \frac{p_h \tilde f_h(Y^\ab \mid Y^\a, Y^\b)}{\sum_{h' \in H} p_{h'} \tilde f_{h'} (Y^\ab \mid Y^\a, Y^\b)}
\end{equation}
where 
\begin{align*}
    \tilde f_h(Y^\ab \mid Y^\a, Y^\b) &= \int \bigg\{\tilde f_h(Y^\ab \mid \tilde \theta_h, \lambda_\a, \lambda_\b) \pi_h(\tilde \theta_h \mid \lambda_\a, \lambda_\b) \times\\
    & ~~~~~~~~~~~~~~~~~~~~~~~~  \pi(\lambda_\a \mid Y^\a) \pi(\lambda^\b \mid Y^\b)\bigg\} d\tilde \theta_h d\lambda_\a d\lambda_\b
\end{align*}
with $\tilde f_h$ denoting the probability mass function of $Y^\ab$ under model $h$ and $\tilde \theta_h$ denoting the remaining parameters of the model. Notice that 
\begin{equation}
\pi(\lambda_\a \mid Y^\a) = \gam(a + S_\a, b + n_\a),\quad\pi(\lambda_\b \mid Y^\b) = \gam(a + S_\b, b + n_\b)
\label{eq:stage1}
\end{equation}
where $S_\a = \sum_{j = 1}^{n_\a} Y^\a_j$ and $S_\b = \sum_{j = 1}^{n_\b} Y^\b_j$. 

A particular integration operation that shows up repeatedly in the following calculations stems from the well-know Poisson-Gamma conjugacy. For a vector of non-negative integers $y = (y_1, \ldots, y_n)$ and positive real numbers $\alpha, \beta$, we define the quantity
\begin{align}
g(y; \alpha, \beta) & := \int \prod_{j = 1}^n \pois(y_j \mid \lambda) \gam(\lambda \mid \alpha, \beta) d\lambda\\ 
& = \frac{\Gamma(\alpha + S(y))}{\Gamma(\alpha)}  \frac{\beta^\alpha}{(\beta + n)^{\alpha + S(y)}} \frac{1}{\prod_{j = 1}^n y_j!}
\end{align}
where $S(y) = \sum_{j = 1}^n y_j$. This quantity could be easily evaluated using any standard mathematics or statistics software. But we do want caution the user of numerical overflow problems as the gamma function grows super-exponentially in its argument. It is best to carry out the calculation of $g(y; \alpha, \beta)$ in the logarithmic scale, such as using the \textsf{lgamma()} function in the R software platform.


\subsubsection{Computation for ``Single'' hypothesis}
Leveraging the Poisson-Gamma conjugacy, one can directly calculate
\begin{align*}
\tilde f(Y^\ab \mid Y^\a, Y^\b) &= \frac12 \bigg[\int \prod_{j = 1}^{n_\ab} \pois(Y^\ab_j \mid \lambda_\a) \pi(\lambda_\a \mid Y^\a) d\lambda_\a\\
&~~~~~~~~~~~~~~~~~~~~ + \int \prod_{j = 1}^{n_\ab} \pois(Y^\ab_j \mid \lambda_\b) \pi(\lambda_\b \mid Y^\b) d\lambda_\b\bigg]\\
& = \frac12 \bigg[g(Y^\ab; a + S_\a, b + n_\a) + g(Y^\ab; a + S_\b, b + n_\b) \bigg].
\end{align*}
The above calculation is done by assuming that the total prior probability of the Single hypothesis is split equally ({\it a priori}) between its two sub-hypotheses. A more conservative variation of this would be to report the maximum of the two numbers $g(Y^\ab; a + S_\a, b + n_\a)$ and $g(Y^\ab; a + S_\b, b + n_\b)$, giving full weight to the sub-hypothesis that explains the data better. Such selective representation of strongest sub-hypothesis has been used in the literature \citep{berger2001bayesian}.

\subsubsection{Computation for ``Mixture'' hypothesis}
For this hypothesis, the remaining parameter is $\tilde \theta = \alpha$ with $\pi(\alpha \mid \lambda_\a, \lambda_\b) = \bet(c_1, c_2)$ and
\begin{equation}
\tilde f(Y^\ab \mid \alpha, \lambda_\a, \lambda_\b) = \prod_{j = 1}^{n_\ab}\{ \alpha \cdot \pois(Y^\ab_j \mid \lambda_\a) + (1 - \alpha) \cdot \pois(Y^\ab_j \mid \lambda_\b)\}. 
\label{eq:mix f}
\end{equation}
This form of $\tilde f$ is difficult to work with directly because when the product of the sum is expanded, it results in too many summands; $2^{n_\ab}$ many to be precise. Instead, a commonly adopted strategy in dealing with mixture model computation is to rewrite the model by introducing latent (unobserved) variables $Z_j \in \{\a,\b\}$, $j = 1, \ldots, n_\ab$ that indicate which of the two Poisson components observation $j$ came from. By considering,
\[
    Z_j \sim \disc(\{\a,\b\}; (\alpha, 1 - \alpha)); ~~
    Y_j \mid (Z_j = c) \sim \pois(\lambda_c); ~~j = 1, \ldots, n_\ab
\]
we recover the same joint distribution $\tilde f$ for $Y^\ab$ as in \eqref{eq:mix f}.

One may write $\tilde f(Y^\ab \mid Y^\a, Y^\b) = \int \tilde f(Y^\ab \mid Y^\a, Y^\b, Z) p(Z) dZ$
where $p(Z)$ denotes the joint distribution on $Z$ under the model (and the integral actually is a sum over a discrete space): $p(Z) = \int p(Z\mid\alpha)\pi(\alpha)d\alpha = B(c_1 + \#\{Z_j = \a\}, c_2 + \#\{Z_j = \b\}) / B(c_1, c_2)$ where $B(\cdot,\cdot)$ is the beta function. The integral can be numerically approximated by importance sampling Monte Carlo as follows. Let $q(Z)$ denote any distribution on the space of $Z$. Then with $Z^m,m = 1, \ldots, M$, denoting a large sample of independent draws of $Z$ from $q(Z)$, one has
\[
\tilde f(Y^\ab \mid Y^\a, Y^\b) \approx \frac1M \sum_{m = 1}^M \tilde f(Y^\ab \mid Y^\a ,Y^\b, Z = Z^m) \frac{p(Z = Z^m)}{q(Z = Z^m)}
\]
by the strong law of large numbers. The quality of this Monte Carlo approximation is improved by choosing an {\it importance distribution} $q(Z)$ that closely resembles the posterior distribution $p(Z \mid Y^\ab, Y^\a, Y^b) \propto \tilde f(Y^\ab \mid Y^\a, Y^\b, Z) p(Z)$; see \citep{tokdar2010importance} for more details. For our purposes, a good and convenient choice is a $q(Z)$ under which $Z_j \sim \disc(\{\a, \b\}, (\bar \alpha_j, 1 - \bar \alpha_j))$, $j = 1, \ldots, n$, where  
\begin{align*}
\bar \alpha_j & = \frac{\int \pois(Y^\ab_j \mid \lambda_\a) \pi(\lambda_\a \mid Y^\a)d\lambda_\a}{\int \pois(Y^\ab_j \mid \lambda_\a) \pi(\lambda_\a \mid Y^\a)d\lambda_\a + \int \pois(Y^\ab_j \mid \lambda_\b) \pi(\lambda_\b \mid Y^\b)d\lambda_\b},\\
& = \frac{g(Y^\ab_j; a + S_\a, b + n_\a)}{g(Y^\ab_j; a + S_\a, b + n_\a) + g(Y^\ab_j; a + S_\b, b + n_\b)},
\end{align*}
which calculates the probability of classifying trial $j$ as having come from condition A, under equal prior odds.

Therefore, to carry out the above importance sampling Monte Carlo, it is sufficient that we evaluate $\tilde f(Y^\ab \mid Y^\a, Y^\b, Z)$. But this can be computed efficiently as
\begin{align*}
    \tilde f(Y^\ab & \mid Y^\a, Y^\b, Z) = \int \tilde f(Y^\ab \mid \lambda^\a, \lambda^\b, Z) \pi(\lambda_\a\mid Y^\a) \pi(\lambda^\b \mid Y^\b) d\lambda_\a d\lambda_\b\\
    & = \left\{\int \prod_{j:Z_j = \a} \pois(Y^\ab_j \mid \lambda_\a) \pi(\lambda_\a \mid Y^\a) d\lambda_\a\right\}
    \times \left\{\int \prod_{j:Z_j = \b} \pois(Y^\ab_j \mid \lambda_\b) \pi(\lambda_\b \mid Y^\b) d\lambda_\b\right\}\\
    & = g(\{Y^\ab_j: Z_j = \a\}; a + S_\a, b + n_\a) \cdot g(\{Y^\ab_j: Z_j = \b\}; a + S_\b, b + n_\b)
\end{align*}
by using the Poisson-Gamma conjugacy.

\subsubsection{Computation for the ``Intermediate'' hypothesis}
For the Intermediate hypothesis, one can use a straight Monte Carlo average to approximate $\tilde f(Y^\ab \mid Y^\a, Y^\b)$ as
\begin{align*}
\tilde f(Y^\ab & \mid Y^\a, y^\b) \\
    & = \int \left\{\int \tilde f(Y^\ab \mid \lambda) \pi(\lambda \mid \lambda_\a, \lambda_\b) d\lambda \right\} \pi(\lambda_\a \mid Y^\a) \pi(\lambda_\b \mid Y^\b)d\lambda_\a d\lambda_\b\\
    & \approx \frac1M \sum_{m = 1}^M \tilde f(Y^\ab \mid \lambda_\a = \lambda^m_\a, \lambda_\b = \lambda^m_\b)
\end{align*}
where $(\lambda^m_\a, \lambda^m_\b)$, $m = 1, \ldots, M$, are independent draws from $\pi(\lambda_\a \mid Y^\a) \times \pi(\lambda_\b \mid Y^\b)$ and, with $\underline{\lambda} = \min(\lambda_\a, \lambda_\b)$, $\overline{\lambda} = \max(\lambda_\a, \lambda_\b)$, $S_\ab = \sum_{j = 1}^{n\ab} Y^\ab_j$,
\begin{align*}
\tilde f(Y^\ab \mid \lambda_\a, \lambda_\b) & = \int \tilde f(Y^\ab \mid \lambda) \pi(\lambda \mid \lambda_\a, \lambda_\b) d\lambda\\
& = \frac{\int_{\underline \lambda}^{\overline \lambda} \prod_{j = 1}^n \pois(Y^\ab_j \mid \lambda) \lambda^{a - 1} e^{-b\lambda} d\lambda}{\int_{\lambdalo}^{\lambdahi} \lambda^{a - 1} e^{-b\lambda} d\lambda}\\
& = \frac{\int_{\lambdalo}^{\lambdahi} \lambda^{a + S_\ab - 1} e^{-(b + n_\ab)\lambda} d\lambda}{\{\prod_{j = 1}^n Y^\ab_j!\}\int_{\lambdalo}^{\lambdahi} \lambda^{a - 1} e^{-b\lambda} d\lambda}\\
& = g(Y^\ab; a, b) \times \frac{F_{a + S_\ab, b + n}(\lambdahi) - F_{a + S_\ab, b + n}(\lambdalo)}{F_{a, b}(\lambdahi) - F_{a, b}(\lambdalo)}
\end{align*}
where $F_{\alpha,\beta}(x)$ is used to denote the cumulative distribution function of $\gam(\alpha,\beta)$.

\subsubsection{Computation for ``Outside'' hypothesis}
Here the computation is done exactly as in the Intermediate hypothesis case, except for the following calculation which accounts for the fact that the conditional prior on $\lambda$ given $\lambda_\a$, $\lambda_\b$ is $\gam(a,b)$ truncated to the complement of the interval $(\lambdalo, \lambdahi)$:
\begin{align*}
\tilde f(Y^\ab \mid \lambda_\a, \lambda_\b) & = g(Y^\ab; a, b) \times \frac{1 - \{F_{a + S_\ab, b + n}(\lambdahi) - F_{a + S_\ab, b + n}(\lambdalo)\}}{ 1- \{F_{a, b}(\lambdahi) - F_{a, b}(\lambdalo)\}}.
\end{align*}

\subsubsection{Additional considerations for non-informative priors}
An actual implementation of the above testing framework requires choosing the hyper-parameters $a, b, c_1, c_2$, all positive valued real numbers. As with any Bayesian analysis, the results will have some dependence on the choice of these hyper-parameters. While expert knowledge about the model animal, brain region and sensory/cognitive task might help to choose reasonable values of these parameters, it may also be desirable to use some {\it default} values that encode minimal prior information about the model parameters. 

One such approach is to use non-informative priors arising from Jeffreys' work. For the prior on the mixing proportion $\alpha$, the Jeffreys prior is $\bet(1/2, 1/2)$ which corresponds to our choice with $c_1 = c_2 = 1/2$. The Jeffreys' prior for estimating the mean of a Poisson distribution is the improper density function $\pi(\mu) \propto 1/\sqrt{\mu}$, $\mu > 0$, which matches, in a limiting sense, our choice of $\gam(a, b)$ with $a = 1/2$ and $b = 0$. This is because the posterior distribution for the Poisson mean under a $\gam(1/2, \beta)$ prior converges to the posterior distribution under the Jeffreys' prior as $\beta \to 0$.

However, such a limiting property does not hold for the marginal likelihood score! In fact, this score is not even well defined under the Jeffreys' prior, since prior density function is defined only up to a multiplicative constant. Also note that the quantity $g(y; \alpha, \beta) \to 0$ as $\beta \to 0$, and, hence working with a small but non-zero $b$ is not an option either, since the resulting marginal likelihood scores for the Intermediate and the Outside hypotheses can be made arbitrarily small by choosing an arbitrarily small $b$.

Such anomalies can be effectively addressed by following the intrinsic Bayes factor approach of \cite{berger1996intrinsic}. The Bayes factor between two hypotheses $h$ and $h'$ is defined as the ratio of the marginal likelihood scores $B_{h,h'}(Y) = f_h(Y) / f_{h'}(Y)$. Notice that
\begin{equation}
    \frac{p_h(Y)}{p_{h'}(Y)} = \frac{p_h}{p_{h'}} \times B_{h,h'}(Y),
    \label{eq:po odds}
\end{equation}
that is the posterior odds between the two hypotheses depends on the data $Y$ only through the Bayes factor. The intrinsic Bayes factor adjustment works for the case where data $Y$ consists of a collection of observations $(Y_1, \ldots, Y_n)$ which, under each hypothesis $h$, are independently distributed with their distributions depending on a parameter $\theta_h \in \Theta_h$, with a prior distribution $\pi_h(\theta_h)$ chosen on $\Theta_h$.

When one or both of $\pi_h(\theta_h)$ and $\pi_{h'}(\theta_{h'})$ are improper, defined only up to an arbitrary scaling factor, \cite{berger1996intrinsic} recommend replacing them with proper distributions  $\pi^\ell_h(\theta_h) = \pi_h(\theta_h \mid Y_\ell)$ and $\pi^\ell_{h'}(\theta_{h'}) = \pi_{h'}(\theta_{h'} \mid Y_\ell)$ obtained by calculating the posterior distribution given a small fraction of the data $Y_\ell = (Y_j : j \in \ell)$, for subset $\ell \subset \{1, \ldots, n\}$ called a training set. A minimal training set is chosen, so that least amount of data is expended in this step of correcting for the impropriety of the prior distributions. Next, one calculates the marginal likelihood scores based on the new priors $\pi_h^\ell(\theta_h)$ and $\pi_{h'}^\ell(\theta_{h'})$, but using only the remaining part of the data $Y \setminus Y_\ell$. The resulting Bayes factor, which depends on the choice of the training set, but does not depend on any arbitrary scaling of the original priors, can be expressed as $B^\ell_{h,h'}(Y) = B_{h,h'}(Y) / B_{h,h'}(Y_\ell)$. To avoid the the effect of the arbitrary choice of the training set, one calculates the intrinsic Bayes factor $\IBF_{h,h'}(Y)$ which is an average of $B^\ell_{h,h'}$ across all minimal training sets $\ell$.

The final averaging could be an arithmetic, geometric or harmonic mean of the training set adjusted Bayes factors. We adopt the geometric mean approach, because it generalizes nicely to the case where one has more than two hypotheses to compare. The geometric mean intrinsic Bayes factor preserves the transitivity property that $\IBF_{h,h''} = \IBF_{h, h'} \times \IBF_{h', h''}$ and conforms to \eqref{eq:po odds} with $B$ replaced with $\IBF$, for every pair of hypotheses $h, h' \in H$. Furthermore, the geometric mean intrinsic Bayes factor approach can be viewed as a direct adjustment to the marginal likelihood score $\IBF_{h,h'}(Y) = f^*_h(Y) / f^*_{h'}(Y)$, where the corresponding intrinsic marginal likelihood score $f^*_h(Y)$ is defined as the geometric mean of $f_h(Y) / f_h(Y_\ell)$ across all minimal training sets $\ell$.

For our four hypotheses, both Outside and Intermediate have improper priors when $b = 0$. For either hypothesis, a single observation is enough to give a proper posterior and hence the minimal training set size is one. Therefore the intrinsic marginal likelihood score adjustment for any or our models is achieved as:
\begin{equation}
\tilde f^*_h(Y^\ab \mid Y^\a, Y^\b) = \frac{\tilde f_h(Y^\ab \mid Y^\a, Y^\b)}{\left[\prod_{\ell = 1}^{n_\ab} \tilde f_h(Y^\ab_\ell \mid Y^\a, Y^\b)\right]^{1/n_\ab}}
\label{eq:if}
\end{equation}
where one uses the formulas derived above for $\tilde f_h(Y^\ab \mid Y^\a, Y^\b)$ with $b \approx 0$. In our implementation we use $b = 10^{-5}$. The adjustments to the Single, Intermediate and Outside hypotheses are straightforward. For the Mixture model, one does not need to run an importance sampling Monte Carlo to calculate the denominator in \eqref{eq:if}. Instead, since for each $\ell \in \{1, \ldots, n\}$ the corresponding $Z_\ell$ has only two possibilities $\{\a, \b\}$ a full enumeration can be done to express the denominator as $(c_1 + c_2)^{-1} [\prod_{\ell = 1}^{n_\ab} (c_1 g(Y^\ab_\ell; a + S_\a, b + n_\a) + c_2 g(Y^\ab_\ell; a + S_\b, b + n_\b)]^{1/n_\ab}$.

\end{document}